\documentclass[preprintnumbers, prd, floatfix, twocolumn, letterpaper, superscriptaddress,nofootinbib]{revtex4-1}

\usepackage{amsfonts}
\usepackage{amsmath}
\usepackage{amssymb,epsf}
\usepackage{latexsym}
\usepackage{graphicx,epsfig}
\usepackage{amssymb}
\usepackage{float}
\usepackage{subfigure}
\usepackage{epstopdf}
\usepackage[colorlinks=true,citecolor=blue,linkcolor=blue,urlcolor=black]{hyperref}
\usepackage{dcolumn}
\usepackage{psfrag}
\usepackage{wrapfig}
\usepackage{makeidx}
\usepackage{epsf}
\usepackage{color}
\usepackage{multirow}
\usepackage{mathtools}

\graphicspath{{Images/}}

\usepackage{tikz}

\definecolor{lime}{HTML}{A6CE39}
\newcommand{\orcidicon}{%
	\begin{tikzpicture}
	\draw[lime, fill=lime] (0,0) 
		circle [radius=0.16] 
		node[white] {{\fontfamily{qag}\selectfont \tiny ID}};
	\draw[white, fill=white] (-0.0625,0.095) 
		circle [radius=0.007];
	\end{tikzpicture}	\hspace{-2mm}
}
\newcommand\orcidJoaoLuis{{\href{https://orcid.org/0000-0003-4148-7372}{\orcidicon}}}
\newcommand\orcidFrancisco{{\href{https://orcid.org/0000-0002-9388-8373}{\orcidicon}}}
\newcommand\orcidGonzalo{{\href{https://orcid.org/0000-0001-9857-0412}{\orcidicon}}}
\begin{document}

\title{Weak-field regime of the generalized hybrid metric-Palatini gravity}

\author{Jo\~{a}o Lu\'{i}s Rosa\orcidJoaoLuis\!\!}
\email{joaoluis92@gmail.com}
\affiliation{Institute of Physics, University of Tartu, W. Ostwaldi 1, 50411 Tartu, Estonia}
\author{Francisco S. N. Lobo\orcidFrancisco\!\!}
\email{fslobo@fc.ul.pt}
\affiliation{Instituto de Astrof\'{i}sica e Ci\^{e}ncias do Espa\c{c}o, Faculdade de Ci\^encias da Universidade de Lisboa, Edif\'{i}cio C8, Campo Grande, P-1749-016, Lisbon, Portugal}
\affiliation{Departamento de F\'{i}sica, Faculdade de Ci\^{e}ncias, Universidade de Lisboa, Edifício C8, Campo Grande, PT1749-016 Lisbon, Portugal}
\author{Gonzalo J. Olmo\orcidGonzalo\!\!} \email{gonzalo.olmo@uv.es}
\affiliation{Departamento de F\'{i}sica Te\'{o}rica and IFIC, Centro Mixto Universidad de Valencia - CSIC.
Universidad de Valencia, Burjassot-46100, Valencia, Spain}
\affiliation{Departamento de F\'isica, Universidade Federal da
Para\'\i ba, 58051-900 Jo\~ao Pessoa, Para\'\i ba, Brazil}

\date{\today}

\begin{abstract} 
In this work we explore the dynamics of the generalized hybrid metric-Palatini theory of gravity in the weak-field, slow-motion regime. We start by introducing the equivalent scalar-tensor representation of the theory, which contains two scalar degrees of freedom, and perform a conformal transformation to the Einstein frame. Linear perturbations of the metric in a Minkowskian background are then studied for the metric and both scalar fields. The effective Newton constant and the PPN parameter $\gamma$ of the theory are extracted after transforming back to the (original) Jordan frame. Two particular cases where the general method ceases to be applicable are approached separately. A comparison of these results with observational constraints is then used to impose bounds on the masses and coupling constants of the scalar fields. 
\end{abstract}


\maketitle

\section{Introduction}\label{sec:intro}

Modified theories of gravity \cite{Nojiri:2010wj,Sotiriou:2008rp,Capozziello:2011et,Avelino:2016lpj,Lobo:2008sg} have recently received much attention, as an alternative to dark energy models \cite{Copeland:2006wr,Horndeski:1974wa,Deffayet:2011gz}, in order to explain the late-time cosmic acceleration \cite{Perlmutter:1998np,Riess:1998cb}. In fact, a popular theory extensively analyzed in the literature is $f(R)$ gravity, which generalizes the Hilbert-Einstein Lagrangian to an arbitrary function of the Ricci curvature scalar $R$. The only restriction imposed on the function $f$ is that it needs to be analytic, namely, it possesses a Taylor expansion about any point. Indeed, earlier interest in $f(R)$ gravity was motivated by inflationary scenarios \cite{Starobinsky:1980te} and it has been extremely successful in accounting for the accelerated expansion of the universe \cite{Capozziello:2002rd,Carroll:2003wy}, where the conditions to have viable cosmological models have also been derived (see \cite{Nojiri:2010wj,Sotiriou:2008rp,Capozziello:2011et} for details). It has also been shown that $f(R)$ gravity is strongly constrained by local observations \cite{Olmo:2006eh,Olmo:2005zr,Olmo:2005hc}, at the laboratory and Solar System scales, unless screening mechanisms are invoked \cite{Brax:2010kv,Brax:2012gr,Brax:2012bsa}.

One may approach $f(R)$ gravity through several formalisms at a fundamental level \cite{Nojiri:2010wj,Sotiriou:2008rp,Capozziello:2011et}, namely, one may consider that the metric represents the fundamental field of the theory, and consequently obtain the gravitational field equations by varying the action with respect to the metric. However, one may also consider the so-called Palatini (metric-affine) formalism \cite{Olmo:2011uz}, where the theory possesses two fundamental fields, namely, the metric and the connection, and the action is varied with respect to both. Note that in general relativity, both metric and Palatini formalisms are equivalent, contrary to $f(R)$ gravity. This is transparent if one considers the scalar-tensor representation of $f(R)$ gravity, where the metric formalism corresponds to a Brans-Dicke type with a parameter $\omega_{\rm BD}=0$, while the Palatini formalism is equivalent to a Brans-Dicke theory with $\omega_{\rm BD}=-3/2$, so that both approaches yield different dynamics.
However, a third approach exists, denoted hybrid metric-Palatini gravity \cite{Harko:2011nh}, that essentially consists of a hybrid combination of both metric and Palatini formalisms, which cures several of the problematic issues that arise in these approaches \cite{Nojiri:2010wj,Sotiriou:2008rp,Capozziello:2011et}.

The linear version of hybrid metric-Palatini gravity consists of adding to the Hilbert-Einstein Lagrangian $R$ an $f({\cal R})$ term constructed {\it a la} Palatini, and it was shown that the theory can pass the Solar System observational constraints even if the scalar field is very light \cite{Harko:2011nh,Capozziello:2015lza,Harko:2018ayt,Harko:2020ibn,Capozziello:2013uya}. This implies the existence of a long-range scalar field, which is able to modify the cosmological \cite{Capozziello:2012ny,Carloni:2015bua} and galactic dynamics \cite{Capozziello:2012qt,Capozziello:2013yha}, but leaves the Solar System unaffected \cite{Capozziello:2013wq}. A plethora of applications exist in the literature, such as in cosmology \cite{Boehmer:2013oxa,Lima:2014aza,Lima:2015nma,Paliathanasis:2020fyp,Rosa:2021ish} and extra-dimensions \cite{Fu:2016szo,Rosa:2020uli}, stringlike configurations \cite{Harko:2020oxq,Bronnikov:2020zob}, black holes and wormholes \cite{Capozziello:2012hr,Bronnikov:2019ugl,Bronnikov:2020vgg,KordZangeneh:2020ixt,Chen:2020evr}, stellar configurations \cite{Danila:2016lqx} and test of binary pulsars \cite{Avdeev:2020jqo}, among other applications (we refer the reader to \cite{Harko:2018ayt,Harko:2020ibn} for more details).
However, one may consider further generalizations of the linear hybrid metric-Palatini theory, by taking into account an $f(R,{\cal R})$ extension \cite{Tamanini:2013ltp,Koivisto:2013kwa}. Further applications have been considered to cosmological models \cite{Rosa:2017jld,Rosa:2019ejh,Luis:2021xay}, and compact objects \cite{Rosa:2018jwp,Rosa:2020uoi}. 

In fact, one can show that the generalized hybrid metric-Palatini theory of gravity admits a scalar-tensor representation in terms of two interacting scalar fields. In this context, it was shown that upon an appropriate choice of the interaction potential, one of the scalar fields behaves like dark energy, inducing a late-time accelerated expansion of the universe, while the other scalar field behaves like pressureless dark matter that, together with ordinary baryonic matter, dominates the intermediate phases of cosmic evolution. It has been argued that this unified description of dark energy and dark matter gives rise to viable cosmological solutions, which reproduces the main features of the evolution of the universe \cite{Sa:2020qfd,Sa:2020fvn,Sa:2021eft}. It is also interesting to note that recently a class of scalar-tensor theories has been proposed that includes non-metricity, so that it unifies the metric, Palatini and hybrid metric-Palatini gravitational actions with a non-minimal interaction \cite{Borowiec:2020lfx}.

It is important to further investigate the nature of the additional scalar degrees of freedom contained in the generalized hybrid metric-Palatini gravity in the weak-field limit. In \cite{Bombacigno:2019did}, it was shown that performing an analysis at the lowest order of the parametrized post-Newtonian structure of the model, one scalar field can have long range interactions, mimicking in that way dark matter effects. In the context of  gravitational waves propagation, it was shown that it is possible to have well-defined physical degrees of freedom, provided by suitable constraints on model parameters.

In this work, we build on the latter work and pursue the analysis of the post-Newtonian corrections in the scalar-tensor representation of the generalized hybrid metric-Palatini gravity in the Einstein frame. Using an adequate redefinition of the scalar fields, we show that one of scalar degrees of freedom of the theory contributes to the enhancement of the gravitational attraction, while the other mediates a repulsive force. These results are consistent and weakly constrained by observations, although a model for which the scalar fields are short-ranged seems to be preferable.

The work is outlined in the following manner: In Sec. \ref{sec:theory}, we present the action and field equations, and the scalar-tensor representation in the Jordan and Einstein frames, of the generalized hybrid metric-Palatini gravity. In Sec. \ref{sec:weak}, we consider in detail the weak field regime and analyze the perturbative field equations around a Minkowski background in the Jordan and Einstein frames, including a few particular cases of interest that must be considered separately. Finally, in Sec. \ref{sec:concl}, we discuss our results and conclude.

\section{Generalized hybrid metric-Palatini gravity}\label{sec:theory}

\subsection{Action and equations of motion}

Consider the action $S$ of the generalized hybrid metric-Palatini gravity given by
\begin{equation}\label{genaction}
S=\frac{1}{2\kappa^2}\int_\Omega\sqrt{-g}f\left(R,\mathcal R\right)d^4x+\int_\Omega\sqrt{-g}\mathcal L_md^4x,
\end{equation}
where $\kappa^2=8\pi G$, $G$ is the gravitational constant, we use units in which the speed of light is $c=1$, $\Omega$ is the spacetime volume, $g$ is the determinant of the spacetime metric $g_{ab}$, where latin indexes $a,b$ run from $0$ to $3$, $R$ is the metric Ricci scalar, $\mathcal R=g^{ab}\mathcal{R}_{ab}$ is the Palatini Ricci scalar, where the Palatini Ricci tensor is written in terms of an independent connection $\hat{\Gamma}^c_{ab}$ as $\mathcal{R}_{ab}=\partial_c\hat\Gamma^c_{ab}-\partial_b\hat\Gamma^c_{ac}+\hat\Gamma^c_{cd}\hat\Gamma^d_{ab}-\hat\Gamma^c_{ad}\hat\Gamma^d_{cb}$, $\partial_a$ denotes a partial derivative with respect to the coordinate $x^a$, $f\left(R,\mathcal R\right)$ is a well-behaved function of $R$ and $\mathcal R$, and $\mathcal L_m$ is the matter Lagrangian minimally coupled to the metric $g_{ab}$.

A variation of Eq. \eqref{genaction} with respect to the metric $g_{ab}$ yields the modified field equations
\begin{eqnarray}
\frac{\partial f}{\partial R}R_{ab}+\frac{\partial f}{\partial \mathcal{R}}\mathcal{R}_{ab}-\frac{1}{2}g_{ab}f\left(R,\cal{R}\right)   \nonumber \\
-\left(\nabla_a\nabla_b-g_{ab}\Box\right)\frac{\partial f}{\partial R}=\kappa^2 T_{ab}\label{genfield},
\end{eqnarray}
where $\nabla_a$ denotes a covariant derivative and $\Box=\nabla^a\nabla_a$ is the d'Alembert operator, both written in terms of the metric $g_{ab}$, and $T_{ab}$ is the stress-energy tensor defined in the usual manner as
\begin{equation}\label{setensor}
T_{ab}=-\frac{2}{\sqrt{-g}}\frac{\delta\left(\sqrt{-g}\mathcal L_m\right)}{\delta\left(g^{ab}\right)}.
\end{equation}

On the other hand, varying Eq. \eqref{genaction} with respect to the independent connection $\hat\Gamma^c_{ab}$, the relevant part of the connection equation can be written as 
\begin{equation}\label{eomgamma}
\hat\nabla_c\left(\sqrt{-g}\frac{\partial f}{\partial\mathcal R}g^{ab}\right)=0,
\end{equation}
where $\hat\nabla_a$ is the covariant derivative written in terms of $\hat\Gamma^c_{ab}$. For a detailed account of the role of torsion in the derivation of the above equation, see \cite{Afonso:2017bxr}. From that result one finds that for bosonic fields, which is the case we are interested in here, torsion can be trivialized via a projective transformation. Standard algebraic manipulations then lead us to conclude  that there exists a metric $\hat{g}_{ab}$ conformally related to $g_{ab}$ defined as
\begin{equation}\label{defhab}
\hat{g}_{ab}=\frac{\partial f}{\partial \mathcal R}g_{ab},
\end{equation}
for which the connection $\hat\Gamma^c_{ab}$ is the Levi-Civita connection, i.e., we can write
\begin{equation}
\hat\Gamma^a_{bc}=\frac{1}{2}\hat{g}^{ad}\left(\partial_b \hat{g}_{dc}+\partial_c \hat{g}_{bd}-\partial_d \hat{g}_{bc}\right).
\end{equation}

\subsection{Scalar-tensor representation}

In a wide variety of cases of interest, it is useful to express the action given in Eq. \eqref{genaction} in a dynamically equivalent scalar-tensor representation. This can be achieved via the addition of two auxiliary fields $\alpha$ and $\beta$ in the following form
\begin{eqnarray}\label{auxaction}
S&=&\frac{1}{2\kappa^2}\int_\Omega\sqrt{-g}\left[f\left(\alpha,\beta\right)+\frac{\partial f}{\partial\alpha}\left(R-\alpha\right)\right.
	\nonumber \\
&&\left.+ \frac{\partial f}{\partial\beta}\left(R-\beta\right)\right]d^4x+\int_\Omega\sqrt{-g}\mathcal L_md^4x.
\end{eqnarray}
At this point one verifies that setting $\alpha=R$ and $\beta=\mathcal R$ recovers the original action \eqref{genaction}. Let us define two scalar fields $\varphi$ and $\psi$ by the following
\begin{equation}
\varphi=\frac{\partial f}{\partial \alpha}, \qquad \psi=\frac{\partial f}{\partial \mathcal \beta} .
\end{equation}

With these definitions, the auxiliary action \eqref{auxaction} takes the form
\begin{eqnarray}\label{almostaction}
S&=&\frac{1}{2\kappa^2}\int_\Omega\sqrt{-g}\left[\varphi R +\psi\mathcal R-V\left(\varphi,\psi\right)\right]d^4x
	\nonumber \\
&&+\int_\Omega\sqrt{-g}\mathcal L_md^4x,
\end{eqnarray}
where the function $V\left(\varphi,\psi\right)$ assumes the role of the scalar fields interaction potential and is defined as
\begin{equation}
V\left(\varphi,\psi\right)=-f\left(\alpha,\beta\right)+\varphi\alpha+\psi\beta,
\end{equation}
and the auxiliary fields $\alpha$ and $\beta$ should be regarded as functions of $\varphi$ and $\psi$. Given the conformal relation between $\hat{g}_{ab}$ and $g_{ab}$ provided in Eq. \eqref{defhab}, which becomes $\hat{g}_{ab}=+\psi g_{ab}$ according to the definitions above, one can show that the $R$ and $\mathcal R$ are related via the expression
\begin{equation}
\mathcal R=R+\frac{3}{\psi^2}\partial^a\psi\partial_a\psi-\frac{3}{\psi}\Box\psi.
\end{equation}
This allows us to eliminate the dependence in $\mathcal R$ of the action given in Eq. \eqref{almostaction}, thus yielding
\begin{eqnarray}
S&=&\frac{1}{2\kappa^2}\int_\Omega\sqrt{-g}\left[\left(\varphi+\psi\right)R+\frac{3}{2\psi}\partial^a\psi\partial_a\psi \right.
	\nonumber\\
&&\left.-V\left(\varphi,\psi\right)\right]d^4x+\int_\Omega\sqrt{-g}\mathcal L_md^4x.\label{jordanaction1}
\end{eqnarray}

The action in Eq. \eqref{jordanaction1} has proven to be useful in numerous analyses. However, in this case we will perform an additional redefinition of the scalar fields for convenience. Consider the scalar fields $\phi$ and $\lambda$ defined as
\begin{equation}
\phi=\varphi+\psi,\qquad s\lambda^2=\psi,
\end{equation}
where $s=\pm 1$ represents the sign of $\psi$. With these definitions, Eq. \eqref{jordanaction1} becomes
\begin{eqnarray}\label{jordanframe}
S&=&\frac{1}{2\kappa^2}\int_\Omega\sqrt{-g}\left[\phi R+6s\partial^a\lambda\partial_a\lambda\right.
	\nonumber\\
&&\left.-\bar V\left(\phi,\lambda\right)\right]d^4x+\int_\Omega\sqrt{-g}\mathcal L_md^4x,
\end{eqnarray}
where $\bar V$ is a new potential written in terms of the scalar fields $\phi$ and $\lambda$. 
The action in Eq. \eqref{jordanframe} describes the scalar-tensor representation of the theory in the Jordan frame. The weak-field phenomenology of the theory in this frame has already been explored in \cite{Bombacigno:2019did}. We shall now perform a change of frame to the Einstein frame to carry out the analysis in those variables.

\subsection{Equations in the Einstein frame}

To switch from the Jordan frame to the Einstein frame, we perform a conformal transformation in the metric of the form $\tilde g_{ab}=\phi g_{ab}$. Consequently, the action in Eq. \eqref{jordanframe} takes the form
\begin{eqnarray}
S&=& \frac{1}{2\kappa^2}\int_\Omega\sqrt{-\tilde g}\left[\tilde R+\frac{6s}{\phi}\tilde \nabla_a\lambda\tilde\nabla^a\lambda-\frac{3}{2\phi^2}\tilde \nabla_a\phi\tilde\nabla^a\phi \right.
	\nonumber \\
&&\left. -\frac{\bar V\left(\phi,\lambda\right)}{\phi^2}\right]d^4x+\int_\Omega\sqrt{-g}\mathcal L_md^4x.\label{almosteinstein}
\end{eqnarray}
To finalize, we perform one further redefinition of the scalar fields as
\begin{equation}\label{defscatilde}
\tilde \phi = \sqrt{\frac{3}{2}}\frac{\log \phi}{\kappa},\quad\quad\tilde\lambda=\frac{\sqrt{6}}{\kappa}\lambda.
\end{equation}
These redefinitions allow us to write the action in the final form 
\begin{eqnarray}\label{finalaction}
S&=&\int_\Omega\sqrt{-\tilde g}\left[\frac{\tilde R}{2\kappa^2}+\frac{s}{2}e^{-\sqrt{\frac{2}{3}}\kappa\tilde\phi}\tilde \nabla_a\tilde\lambda\tilde\nabla^a\tilde\lambda-\frac{1}{2}\tilde \nabla_a\tilde\phi\tilde\nabla^a\tilde\phi \right.
	\nonumber\\
&&\left. -\tilde W\left(\tilde\phi,\tilde\lambda\right)\right]d^4x+\int_\Omega\sqrt{-g}\mathcal L_md^4x,
\end{eqnarray}
with the new potential $\tilde W\left(\tilde\phi,\tilde\lambda\right)$ defined as
\begin{equation}\label{defW}
\tilde W\left(\tilde\phi,\tilde\lambda\right)=\frac{\bar V\left(\phi,\lambda\right)}{2\kappa^2}e^{-2\sqrt{\frac{2}{3}}\kappa\tilde\phi},
\end{equation}
where $\phi$ and $\lambda$ can be written in terms of $\tilde\phi$ and $\tilde\lambda$ via the definitions in Eq.\eqref{defscatilde}. From this point onward, all variables defined in the Einstein frame will be labelled with a tilde. For consistency, we will also denote $\tilde V\left(\tilde\phi,\tilde\lambda\right)\equiv \bar V\left(\phi(\tilde\phi),\lambda(\tilde\lambda)\right)$. 

The action in Eq. \eqref{finalaction} depends on three independent variables, namely the metric $g_{ab}$, and the scalar fields $\phi$ and $\lambda$. Performing a variation of Eq. \eqref{finalaction} with respect to these variables yields, respectively
\begin{eqnarray}\label{eommetric}
\tilde G_{ab} &+& \frac{1}{2}\tilde g_{ab}\left[e^{-2\sqrt{\frac{2}{3}}\kappa\tilde \phi}\tilde V\left(\tilde \phi,\tilde \lambda\right)\right.
	\nonumber \\
&+&\left.\kappa^2\left(\partial_c\tilde \phi\partial^c\tilde \phi-e^{-\sqrt{\frac{2}{3}}\kappa\tilde \phi}s\partial_c\tilde \lambda\partial^c\tilde \lambda\right)\right]
-\kappa^2\partial_a\tilde \phi\partial_b\tilde \phi
	\nonumber\\
&+& \kappa^2se^{-\sqrt{\frac{2}{3}}\kappa\tilde \phi}\partial_a\tilde \lambda\partial_b\tilde \lambda
=\kappa^2e^{-\sqrt{\frac{2}{3}}\kappa\tilde \phi}T_{ab},
\end{eqnarray}
\begin{eqnarray}\label{eomtildephi}
\Box\tilde \phi-\frac{1}{2\kappa^2}e^{-2\sqrt{\frac{2}{3}}\kappa\tilde \phi}\left(\tilde V_{\tilde\phi}-2\sqrt{\frac{2}{3}}\kappa\tilde \phi \ \tilde V\left(\tilde \phi,\tilde \lambda\right)\right)
	\nonumber\\
-\frac{1}{\sqrt{6}}e^{-\sqrt{\frac{2}{3}}\kappa\tilde \phi}s\kappa\partial_a\tilde \lambda\partial^a\tilde \lambda=\sqrt{\frac{2}{3}}\kappa T,
\end{eqnarray}
\begin{equation}\label{eomtildelambda}
\Box\tilde \lambda-\sqrt{\frac{2}{3}}\kappa\partial_a\tilde \phi\partial^a\tilde \lambda+\frac{s}{2\kappa^2}e^{-\sqrt{\frac{2}{3}}\kappa\tilde \phi}\ \tilde V_{\tilde \lambda}=0,
\end{equation}
where the subscripts $\tilde \phi$ and $\tilde \lambda$ denote partial derivatives with respect to these scalar fields, and $T=\tilde g^{ab}T_{ab}$ is the trace of the stress-energy tensor.

From the above equations, it is worth noting that the scalar field $\tilde\phi$ is sourced by both $\tilde\lambda$ and the matter stress-energy density $T$, whereas $\tilde\lambda$ only couples to itself and to $\tilde\phi$. According to this, $\tilde \lambda$ can be regarded as a kind of dark matter fluid, which gravitates but does not directly feel the presence of matter,  in interaction with the dark energy field $\tilde \phi$. This structure of the field equations suggests potential applications of this type of models to scenarios with interacting dark sectors. 

\section{The weak field regime}\label{sec:weak}

\subsection{Perturbative equations}

Let us now analyze the effects of the scalar fields $\tilde \phi$ and $\tilde \lambda$ in a slightly curved space. To do so, we shall consider a system of local coordinates in which the metric can be written in terms of a Minkowskian spacetime $\tilde \eta_{ab}$ plus a small perturbation $\tilde h_{ab}$
\begin{equation}\label{gabpert}
\tilde g_{ab}\approx\tilde \eta_{ab}+\tilde h_{ab},
\end{equation}
with $|\tilde h_{ab}|\ll 1$. In the same way, the scalar fields will be written as
\begin{equation}\label{fieldpert}
\tilde \phi = \tilde \phi_0+\delta \tilde \phi, \qquad \tilde \lambda = \tilde \lambda_0+\delta\tilde \lambda,
\end{equation}
where $\tilde \phi_0$ and $\tilde \lambda_0$ represent the (approximately constant) background values and $\delta\tilde \phi$ and $\delta\tilde \lambda$ are local fluctuations of order $\mathcal O\left(\tilde h_{ab}\right)$. Note that these fluctuations should vanish outside the region where the metric is described by Eq. \eqref{gabpert}. More relevant, perhaps, is the fact that we have freedom to set $\tilde{\phi}_0$ to zero without loss of generality. This is so because we can choose the constant $\kappa^2$ in Eq. \eqref{jordanframe} such that $\phi_0=1$ at our cosmic reference time $t_0$, thus implying that $\tilde{\phi}_0=0$ according to Eq. \eqref{defscatilde}. For generality, however, we will keep this quantity arbitrary until it becomes convenient to fix its reference value.

In the weak-field regime, derivatives of the background fields are negligible, as the evolution of the scalar fields is very slow due to the large difference between cosmological and solar system scales. Consequently, curvature terms and first order derivatives of the background metric can be discarded. Time derivatives shall also be neglected because the motion of the sources is expected to be non-relativistic, and thus the D'Alembert operator $\Box$ effectively becomes the Laplacian operator $\nabla^2$. Furthermore, we shall assume that matter perturbations are described by a presureless perfect fluid, i.e., we write the perturbed stress-energy tensor $\delta T_{ab}$ as
\begin{equation}
\delta T^{ab}=\rho u^au^b,
\end{equation}
where $\rho$ is the energy density and $u^a$ is the 4-velocity of the fluid elements. This implies that the only non-vanishing component of $\delta T_{ab}$ is $\delta T_{00}=\rho$, with the space components $\delta T_{ij}=0$ vanishing, where the indexes $i, j$ run from $1$ to $3$. Also, the trace becomes $\delta T=-\rho$. Fixing the gauge as 
\begin{equation}
\partial_b\left(\tilde h_a^b-\frac{1}{2}\delta_a^b\tilde h\right)=0 ,
\end{equation}
the resultant equations of motion for the perturbed metric $\tilde h_{ab}$ and the scalar field fluctuations $\delta\tilde \phi$ and $\delta \tilde \lambda$ become,
\begin{eqnarray}\label{eqmetricpert}
&&-\frac{\nabla^2 \tilde h_{ab}}{2}=\kappa^2e^{-\sqrt{\frac{2}{3}}\kappa\tilde \phi_0}\left(\delta T_{ab}-\tilde \eta_{ab}\frac{\delta T}{2}\right)
	\nonumber \\ 
&&\qquad 	- \frac{1}{2}\tilde V e^{-2\sqrt{\frac{2}{3}}\kappa\tilde \phi_0}\left(\tilde h_{ab}-\tilde \eta_{ab}\frac{\tilde h}{2}\right)  
	+\frac{ e^{-2\sqrt{\frac{2}{3}}\kappa\tilde \phi_0}}{6}\tilde \eta_{ab} \times
	\nonumber \\ 
&& \qquad \qquad	 \times \left[\left(3\tilde V_{\tilde \phi}-2\sqrt{6}\kappa \tilde V\right)\delta \tilde \phi+3\tilde V_{\tilde \lambda} \delta \tilde \lambda\right],
\end{eqnarray}
\begin{equation}\label{eqphipert}
\left(\nabla^2-m_\phi^2\right)\delta\tilde \phi=a_{\phi}\delta\tilde \lambda-\sqrt{\frac{2}{3}}\kappa\rho ,
\end{equation}
\begin{equation}\label{eqlambdapert}
\left(\nabla^2-m_\lambda^2\right)\delta\tilde \lambda=a_{\lambda}\delta\tilde \phi ,
\end{equation}
where $m_\phi$ and $m_\lambda$ are the masses of the scalar fields $\delta\tilde \phi$ and $\delta\tilde \lambda$, respectively, and $a_{\phi}$ and $a_{\lambda}$ are the coupling constants between $\tilde \phi$ and $\tilde \lambda$. These quantities can be written in terms of the potential $\tilde W$ and its derivatives as
\begin{equation}
m_\phi^2 
= \frac{1}{2\kappa^2}\left(e^{-2\sqrt{\frac{2}{3}}\kappa\tilde \phi}\tilde V\right)_{\tilde \phi\tilde \phi}  \ , \qquad m_\lambda^2=\frac{1}{2\kappa^2}\tilde V_{\tilde \lambda\tilde \lambda},
\end{equation}
\begin{equation}\label{eq:aphi}
a_{\phi}=
 \frac{1}{2\kappa^2}\left(e^{-2\sqrt{\frac{2}{3}}\kappa\tilde \phi}\tilde V\right)_{\tilde \phi\tilde \lambda} 
\end{equation}
\begin{equation}
a_{\lambda}=
-\frac{{s}}{2\kappa^2}\left(e^{-\sqrt{\frac{2}{3}}\kappa\tilde \phi}\tilde V\right)_{\tilde \lambda\tilde \phi} \,,
\end{equation}
respectively.

\subsection{Analysis of the general case}

\subsubsection{Perturbation equations for the scalar fields}

Equations \eqref{eqphipert} and \eqref{eqlambdapert} constitute a system of two coupled differential equations for $\delta\tilde \phi$ and $\delta\tilde \lambda$. To simplify the analysis, it is useful to perform a new change of variables in order to decouple this system, as was done in \cite{Bombacigno:2019did}. To do so, we write the system of Eqs. \eqref{eqphipert} and \eqref{eqlambdapert} in the following matrix form
\begin{equation}\label{mateq1}
\left(I_{2\times 2}\nabla^2-A\right)\Phi=\mathcal T,
\end{equation}
where $I_{2\times 2}$ is the identity matrix in two dimensions and we define the matrix $A$ by
\begin{equation}\label{defA}
A=\begin{pmatrix}m_\lambda^2 & a_{\lambda} \\ a_{\phi} & m_\phi^2\end{pmatrix},
\end{equation}
and the vectors $\Phi$ and $\mathcal T$ as
\begin{equation}
\Phi=\begin{pmatrix}\delta\tilde \lambda \\ \delta\tilde \phi\end{pmatrix}, \qquad
\mathcal T=-\sqrt{\frac{2}{3}}\kappa\begin{pmatrix}\rho \\ 0\end{pmatrix},
\end{equation}
respectively.

A decoupled system of equations can be obtained via the diagonalization of Eq. \eqref{mateq1}. Let $P$ be the matrix of the eigenvectors of $A$ and $P^{-1}$ its inverse. These two matrices take the forms
\begin{equation}\label{defP}
P=\begin{pmatrix}p_{11} & p_{12} \\ p_{21} & p_{22}\end{pmatrix}=\begin{pmatrix}-\frac{M_+^2-m_\lambda^2}{a_\phi} & \frac{M_+^2-m_\phi^2}{a_\phi} \\ 1 & 1\end{pmatrix},
\end{equation}
\begin{equation}\label{defPinv}
P^{-1}=\begin{pmatrix}\bar p_{11} & \bar p_{12} \\ \bar p_{21} & \bar p_{22}\end{pmatrix}=\begin{pmatrix}-\frac{a_\phi}{M^2_0} & \frac{M_+^2-m_\phi^2}{M^2_0} \\ \frac{a_\phi}{M^2_0} & \frac{M_+^2-m_\lambda^2}{M^2_0}\end{pmatrix},
\end{equation}
where we have defined the auxiliary constants $M_\pm^2$ and $M_0$ (with units of mass) as the combinations
\begin{equation}\label{eq:mpm}
M_\pm^2=\frac{1}{2}\left[m_\lambda^2+m_\phi^2	\pm M_0^2\right],
\end{equation}
\begin{equation}\label{eq:M0}
M_0^2=\sqrt{4 a_{\lambda}a_{\phi}+\left(m_\lambda^2-m_\phi^2\right)^2}.
\end{equation}
With the forms of $P$ and $P^{-1}$ defined above, the matrix $A_D=P^{-1}AP$ is diagonal. Let us also define the new scalar field vector as $\Phi_D=P^{-1}\Phi$ and the new matter vector as $\mathcal T_D=P^{-1}\mathcal T$. As a result, Eq. \eqref{mateq1} becomes
\begin{equation}
\left(I_{2\times 2}\nabla^2-A_D\right)\Phi_D=\mathcal T_D.
\end{equation}
The decoupled version of the system of Eqs. \eqref{eqphipert} and \eqref{eqlambdapert} becomes then
\begin{equation}\label{eqphidec}
\left(\nabla^2-M_\phi^2\right)\delta\phi_D=-\bar p_{21}\sqrt{\frac{2}{3}}\kappa\rho,
\end{equation}
\begin{equation}\label{eqlambdadec}
\left(\nabla^2-M_\lambda^2\right)\delta\lambda_D=-\bar p_{11}\sqrt{\frac{2}{3}}\kappa\rho,
\end{equation}
where $\delta\phi_D$ and $\delta\lambda_D$ are the decoupled scalar fields, and $M_\phi$ and $M_\lambda$ are their respective masses. The new scalar fields and masses can be written in terms of the old scalar fields $\delta\phi$ and $\delta\lambda$, as well as their masses $m_\phi$ and $m_\lambda$, and their coupling constants $a_{\phi}$ and $a_{\lambda}$, as well as the previously defined constants $M_\pm^2$ and $M^2_0$ as
\begin{equation}
\delta\phi_D=\frac{1}{M_0^2}\left[\left(M_+^2-m_\lambda^2\right)\delta\tilde \phi+a_{\phi}\delta\tilde \lambda\right],
\end{equation}
\begin{equation}
\delta\lambda_D=\frac{1}{M_0^2}\left[\left(M_+^2-m_\phi^2\right)\delta\tilde \phi-a_{\phi}\delta\tilde \lambda\right],
\end{equation}
\begin{equation}\label{sfDmasses}
M_\phi^2=M_-^2, \qquad M_\lambda^2=M_+^2.
\end{equation}

We are now able to solve Eqs. \eqref{eqphidec} and \eqref{eqlambdadec} with the usual Laplace transform methods, i.e., we write both $\delta\lambda_D$ and $\delta\phi_D$ in terms of their Laplace transforms $\delta\tilde\lambda_D$ and $\delta\tilde\phi_D$, respectively, insert these forms into Eqs. \eqref{eqphidec} and \eqref{eqlambdadec}, manipulate the results in the momentum space, and invert the Laplace transforms using a convolution. In the end, we arrive at the following solutions for $\delta\lambda_D$ and $\delta\phi_D$:
\begin{equation}\label{soldlambdaD}
\delta\lambda_D\left(x\right)=\frac{\kappa}{4\pi}\bar p_{11}\sqrt{\frac{2}{3}}\int\frac{\rho\left(x'\right)}{|x-x'|}e^{-M_\lambda|x-x'|}d^3x',
\end{equation}
\begin{equation}\label{soldphiD}
\delta\phi_D\left(x\right)=\frac{\kappa}{4\pi}\bar p_{21}\sqrt{\frac{2}{3}}\int\frac{\rho\left(x'\right)}{|x-x'|}e^{-M_\phi|x-x'|}d^3x'.
\end{equation}

\subsubsection{Perturbation equations for the metric}

Let us now turn to the metric equations given in Eq. \eqref{eqmetricpert}. The second term on the RHS is proportional to the potential $\tilde W$, which is assumed to be of the order of the cosmological constant. In the weak-field, slow-motion regime used for solar system tests, the contributions of the potential are thus negligible when compared to the local sources given by the stress-energy tensor of the fluid contribution. Thus, we shall neglect this term. Finally, the last term on the RHS depends on products between potential terms and perturbations in the scalar fields. In the Einstein frame, these terms are of the order of magnitude of those coming from scalar fields in the matter sector, which are also negligible when compared to the dominant fluid terms. Therefore, these terms shall also be discarded (which justifies the absence of any dependence on the sign $s$ on the right-hand side of the metric perturbation equations). Consequently, the two independent equations for the metric take the forms
\begin{equation}
\nabla^2\tilde h_{00}=-\kappa^2 e^{-\sqrt{\frac{2}{3}}\kappa\tilde \phi_0}\rho,
\end{equation}
\begin{equation}
\nabla^2\tilde h_{ij}=-\delta_{ij}\kappa^2 e^{-\sqrt{\frac{2}{3}}\kappa\tilde \phi_0}\rho,
\end{equation}
where $\delta_{ij}$ is the Kronecker delta. The above equations can be integrated directly to yield the following solutions for $\tilde h_{00}$ and $\tilde h_{ij}$
\begin{equation}\label{solh00tilde}
\tilde h_{00}=\frac{\kappa^2}{4\pi}e^{-\sqrt{\frac{2}{3}}\kappa\tilde \phi_0}\int\frac{\rho\left(x'\right)}{|x-x'|}d^3x',
\end{equation}
\begin{equation}\label{solhijtilde}
\tilde h_{ij}=\delta_{ij}\frac{\kappa^2}{4\pi}e^{-\sqrt{\frac{2}{3}}\kappa\tilde \phi_0}\int\frac{\rho\left(x'\right)}{|x-x'|}d^3x'.
\end{equation}

From Eqs. \eqref{solh00tilde} and \eqref{solhijtilde} one can extract the PPN parameters of the theory in the Einstein frame. To do so, recall that we have considered a system of units where $\kappa^2=8\pi G$ and that the integral in these equations represents the Newtonian potential. This way, we can write
\begin{eqnarray}
\tilde h_{00} &=& 2\tilde G_{\text{EF}}U_N\left(x\right), \label{ppn1}
	 \\
\tilde h_{ij} &=& 2\delta_{ij}\tilde G_{\text{EF}}\tilde \gamma U_N\left(x\right),
\label{ppn2}
\end{eqnarray}
where $\tilde G_{\text{EF}}$ is the effective gravitational constant in the Einstein frame, $\tilde \gamma$ is a PPN parameter, $U_N\left(x\right)$ is the Newtonian potential written in terms of the distance to the source $x$. Thus, we verify that
\begin{equation}\label{defgeff}
\tilde \gamma=1,\qquad \tilde G_{\text{EF}}=Ge^{-\sqrt{\frac{2}{3}}\kappa\tilde \phi_0}=G/\phi_0.
\end{equation}
Consequently, we observe that in the Einstein frame, the parameter $\tilde \gamma$ is the same as in GR and that the effective gravitational constant $\tilde G_{\text{EF}}$ is a simple rescaling of the Newtonian constant $G$ depending on the background field $ \phi_0$.

\subsubsection{Recovering the results in the Jordan frame}\label{sec:backframe}

Let us now perform the inverse conformal transformation back to the Jordan frame in such a way that we can compare our results to the ones previously obtained in \cite{Bombacigno:2019did}. To do so, let us start by solving the integrals in Eqs. \eqref{soldlambdaD}, \eqref{soldphiD}, \eqref{solh00tilde} and \eqref{solhijtilde} far from a spherically symmetric source. The solutions take the forms
\begin{equation}
\delta\lambda_D\left(r\right)=\frac{\kappa}{4\pi}\bar p_{11}\sqrt{\frac{2}{3}}\frac{M_\odot}{r}e^{-M_\lambda r},
\end{equation}
\begin{equation}
\delta\phi_D\left(r\right)=\frac{\kappa}{4\pi}\bar p_{21}\sqrt{\frac{2}{3}}\frac{M_\odot}{r}e^{-M_\phi r},
\end{equation}
\begin{equation}
\tilde h_{00}\left(r\right)=\frac{2\tilde G_{\text{EF}}M_\odot}{r},
\end{equation}
\begin{equation}
\tilde h_{ij}\left(r\right)=\delta_{ij}\frac{2\tilde G_{\text{EF}}M_\odot}{r},
\end{equation}
where $M_\odot$ is the mass of the source and $r$ is the radial distance from the source. Note that to perform the inverse conformal transformation, we only care about the scalar field $\tilde \phi$, as the scalar field $\tilde \lambda$ was not involved in the transformation. Thus, let us use $\Phi=P\Phi_D$ to recover $\delta\tilde \phi$ in terms of $\delta\lambda_D$ and $\delta\phi_D$ as $\delta\tilde \phi=p_{21}\delta\lambda_D+p_{22}\delta\phi_D$, or more explicitly 
\begin{equation}
\delta\tilde\phi\left(r\right)=\sqrt{\frac{2}{3}}\frac{\kappa M_\odot}{4\pi r}\left(p_{21}\bar p_{11}e^{-M_\lambda r}+p_{22}\bar p_{21}e^{-M_\phi r}\right).
\end{equation}

To recover the solutions in the Jordan frame, we need to find the scalar field $\phi$ used for the conformal transformation. This field is related to the field $\tilde\phi$ as written in Eq. \eqref{defscatilde}. Inserting the relations $\tilde\phi=\tilde\phi_0+\delta\tilde\phi$ and $\phi=\phi_0+\delta\phi$ and keeping only the terms up to first order, we verify that 
\begin{equation}\label{phisrel}
\delta\tilde\phi=\sqrt{\frac{3}{2}}\frac{\delta\phi}{\phi_0\kappa},
\end{equation}
which allows us to write $\delta\phi$ in the form
\begin{eqnarray} \label{eq:deltaphi}
\delta\phi\left(r\right)&=&\frac{4 G M_\odot}{3r}\left(p_{21}\bar p_{11}e^{-M_\lambda r}+p_{22}\bar p_{21}e^{-M_\phi r}\right) \nonumber \\
&=& \frac{4 G M_\odot}{3r}\frac{a_\phi}{M^2_0}\left(e^{-M_\phi r}-e^{-M_\lambda r}\right) .
\end{eqnarray}

We are now in conditions to perform the inverse conformal transformation. At this point one should recall that, by a convenient choice of units, we had freedom to set $\phi_0=1$ at the reference cosmic time $t_0$, in such a way that the perturbation of $\tilde g_{ab}=\phi g_{ab}$ yields a consistent zeroth order Minkowskian limit in the coordinates chosen. Accordingly, using the expansions $\tilde g_{ab}=\tilde\eta_{ab}+\tilde h_{ab}$ and $g_{ab}=\eta_{ab}+ h_{ab}$ for the metrics, $\phi=1+\delta\phi$ for the scalar field, and keeping only the first order terms, we obtain
\begin{equation}\label{habsrel}
\tilde h_{ab}=h_{ab}+\eta_{ab}\delta\phi.
\end{equation}
This result allows us to write the solutions for the perturbations $h_{00}$ and $h_{ij}$ in the forms
\begin{equation}
h_{00}\left(r\right)=\frac{2GM_\odot}{r}\left[1+\frac{2}{3}\frac{a_\phi}{M^2_0}\left(e^{-M_\phi r}-e^{-M_\lambda r}\right)\right],
\end{equation}
\begin{equation}
h_{ij}\left(r\right)=\delta_{ij}\frac{2GM_\odot}{r}\left[1-\frac{2}{3}\frac{a_\phi}{M^2_0}\left(e^{-M_\phi r}-e^{-M_\lambda r}\right)\right],
\end{equation}
where we have used Eqs. \eqref{defgeff}  and \eqref{phisrel} to write $\tilde G_{\text{eff}}=G/\phi_0 $, from which we can extract the effective gravitational constant in the Jordan frame $G_{\text{eff}}$ and the $\gamma$ PPN parameter given by
\begin{equation}\label{Geff1}
G_{\text{eff}}=G\left[1+\frac{2}{3}\frac{a_\phi}{M^2_0}\left(e^{-M_\phi r}-e^{-M_\lambda r}\right)\right],
\end{equation}
\begin{equation}\label{gammaeff1}
\gamma=\frac{3M^2_0-2a_\phi\left(e^{-M_\phi r}-e^{-M_\lambda r}\right)}{3M^2_0+2a_\phi \left(e^{-M_\phi r}-e^{-M_\lambda r}\right)} ,
\end{equation}
respectively. At this stage, we confirm that the exponential dependences of $h_{00}$, $h_{ij}$ and $\delta\phi$ presented here are consistent with those found in 
\cite{Bombacigno:2019did} [see their equations (50-52) and (55-56)], though there is no transparent correspondence between parameters due to the various redefinitions and assumptions involved. In any case, assuming that the effective potentials are at a minimum, our definition of $M_\phi$ coincides with theirs, and our $M_\lambda$ corresponds to their $M_\xi$.

From the above results one readily sees that the scalar degrees of freedom contribute in a mixed manner to $G_{\text{eff}}$, with a piece that enhances the gravitational attraction (proportional to $e^{-M_\phi r}$) and another that mediates a repulsive force (proportional to $e^{-M_\lambda r}$). It is tempting to argue that the existence of this repulsive force could have been guessed already from the action in Eq. \eqref{jordanaction1}, where the kinetic term associated to $\psi$ appears with a {\it positive} sign. Indeed, the transition to the representation in terms of $\lambda$ took care of this fact by specifying the possibility of splitting the domain of $\psi$ in two sectors with different signs, in such a way that for $s=+1$ the action in Eq. \eqref{jordanframe} can be seen as representing a ghost scalar $\lambda$ while for $s=-1$ it contributes with a positive kinetic energy. In this latter case, one should make sure that the combination $\phi=\varphi+\psi$ (with $\psi<0$) does not change sign in non-perturbative scenarios in order to avoid breakdowns in the evolution of initial data. However, it should also be noted that the sign $s$ enters in the expressions for $G_{\text{eff}}$ and $\gamma$ in a non-linear manner, via the definitions of $M_\phi^2$ and $M_\lambda^2$. The repulsive character of the $e^{-M_\lambda r}$ term, therefore, cannot be directly related to the sign of $s$ but rather to some nontrivial combination of the two dynamical scalar degrees of freedom present in the theory. 

Compatibility with observations requires that the radial dependence of $G_{\text{eff}}$ be negligible within the scales accessible to observations. This can be achieved in different ways. One of them is by making the amplitude $a_\phi/M^2_0$ sufficiently small. Another possibility would be to have very massive scalar modes, such that $M_\phi r$ and $M_\lambda r$ become much bigger than unity, leading to vanishing exponentials. Both possibilities would automatically recover the predictions of GR. A third possibility is to have very light fields, with $M_\phi r$ and $M_\lambda r$ approaching zero in the scales of interest. Assuming that these products are small and expanding the exponentials as $e^{-M r}\approx 1-Mr +O(M^2 r^2)$, we find that 
\begin{equation}\label{eq:gammaeffapp1}
\gamma\approx 1-\frac{4a_\phi}{3M^2_0}\left(M_\lambda-M_\phi\right)L \ ,
\end{equation}
with $L$ being a scale of the order of a few astronomical units. Given that current data set $|\gamma-1|<10^{-5}$, it follows that $|\frac{4a_\phi}{3M^2_0}\left(M_\lambda-M_\phi\right)L|<10^{-5}$, which sets a weak constraint on the model parameters.  

The limit in which $M_\phi^2$ becomes degenerate with $M_\lambda^2$ deserves some attention because it coincides with $M_0^2\to 0$. Taking the limit $M_0^2\to 0$ in Eqs. \eqref{Geff1} and \eqref{gammaeff1}, one finds that this limit is smooth, leading to 
\begin{equation}\label{Geff2}
G_{\text{eff}}=G\phi_0\left[1+\frac{\sqrt{2}}{3}\frac{a_\phi}{M}re^{-\frac{M r}{\sqrt{2}}}\right],
\end{equation}
\begin{equation}\label{gammaeff2}
\gamma=\frac{3M-\sqrt{2}a_\phi r e^{-\frac{M r}{\sqrt{2}}}}{3M+\sqrt{2}a_\phi r e^{-\frac{M r}{\sqrt{2}}}}.
\end{equation}
As we can see, in this limit the repulsive component in the effective Newton's constant disappears, and compatibility with experiments still requires a short range field or a small amplitude $a_\phi/M$, where we have defined $M^2\equiv  m_\lambda^2+m_\phi^2$. An expansion similar to Eq. \eqref{eq:gammaeffapp1} then leads to 
\begin{equation}\label{eq:gammaeffapp2}
\gamma\approx 1-\frac{2\sqrt{2}a_\phi}{3M}Le^{-\frac{M L}{\sqrt{2}}} \ ,
\end{equation}
which also sets a weak constraint on the parameters.

It should  be noted that the particular case $a_\phi=0$ must be analyzed independently and cannot be guessed from these general formulas given above. We also find troubles when the matrix $A$ in Eq. \eqref{defA} becomes degenerate, which forces a reconsideration of the method used to solve the equations. These two particular cases will be studied next.

\subsection{Non-diagonalizable matrix $A$ \label{sec:C}}

The approach presented in the previous section can only be applied in the general case where the matrix A, given by Eq. \eqref{defA}, is diagonalizable. In fact, if one considers the particular case for which the determinant of the matrix $A$ vanishes, it follows that $P$ in Eq.\eqref{defP} becomes a matrix of rank 1 and ceases to be invertible. As a consequence, $A$ will only have one eigenvalue with algebraic degeneracy of $2$ and only one eigenvector, which confirms that in this case it is not diagonalizable anymore and a different method is necessary to solve the system of equations. The condition that the determinant of the matrix $A$ vanishes is equivalent to the relation $m_\phi^2 m_\lambda^2-a_\phi a_\lambda=0$ and forces a separate analysis of that particular case.

Performing a Fourier transform of Eqs. \eqref{eqphipert} and \eqref{eqlambdapert} we find the following relation
\begin{equation}
\delta\hat{\tilde \phi}=\kappa\sqrt{\frac{2}{3}}\frac{(k^2+m_\lambda^2)}{k^2(k^2+M^2)}\hat\rho \ ,
\end{equation}
where we used $M^2\equiv m_\phi^2+m_\lambda^2$ (already defined when considering the limit $M_0^2\to 0$), and a hat denotes a Fourier transform. Considering that $\rho(\vec{x})$ represents a delta-like distribution, $ \rho(\vec{x}')=M_\odot \delta^{(3)}(\vec{x}')$, to simplify the integrations, we find that 
\begin{equation}
\delta\tilde\phi(r)=\kappa \sqrt{\frac{2}{3}}\frac{M_\odot}{4\pi M^2 r}\left(m_\lambda^2+m_\phi^2 e^{-M r}\right) \ .
\end{equation}

On the other hand, the expression for $\delta\hat {\tilde\lambda}$ becomes, 
\begin{equation}
\delta\hat{\tilde \lambda}=-\frac{a_\lambda \delta\hat{\tilde \phi}}{(k^2+m_\lambda^2)} \ ,
\end{equation}
which after some algebraic manipulations yields
 \begin{equation}
\delta\tilde\lambda(r)=-a_\lambda\sqrt{\frac{2}{3}}\frac{\kappa M_\odot}{4\pi M^2 r}\left(1-e^{-M r}\right) \ .
\end{equation}
Consequently, inverting the conformal transformation and using Eq. \eqref{habsrel}, the metric perturbations can be found to be
\begin{eqnarray}
h_{00}&=&\frac{2GM_\odot}{r}\left[1+\frac{2}{3M^2}\left(m_\lambda^2+m_\phi^2e^{-M r}\right)\right] \,, \\
h_{ij}&=&\frac{2GM_\odot}{r}\left[1-\frac{2}{3M^2}\left(m_\lambda^2+m_\phi^2e^{-M r}\right)\right]\delta_{ij} \ .
\end{eqnarray}

Using the definitions in Eqs. \eqref{ppn1} and \eqref{ppn2}, one can again extract both the effective gravitational constant $G_{\text{eff}}$ and the $\gamma$ PPN parameter, which in this case are given by
\begin{eqnarray}
G_{\text{eff}}&=&G\left[1+\frac{2}{3M^2}\left(m_\lambda^2+m_\phi^2e^{-M r}\right)\right] \,, \\ 
\gamma_{\text{eff}}&=&\frac{3M^2-2\left(m_\lambda^2+m_\phi^2e^{-M r}\right)}{3M^2+2\left(m_\lambda^2+m_\phi^2e^{-M r}\right)} \,,
\end{eqnarray}
respectively.
The expression for $G_{\text{eff}}$ indicates that the repulsive degree of freedom mediated by a combination of the two scalar fields in the general case is no longer present when the determinant of $A$ vanishes. The net effect on $G_{\text{eff}}$ is a constant shift of its bare value plus a standard (attractive) Yukawa-type correction. In a sense, we could say that one of the resulting scalar degrees of freedom has infinite range (vanishing mass) while the other has a range $1/M$. This is consistent with the fact that for this choice of parameters $M_0^2$ becomes $M_0^2=M^2$ and leads to $M_+^2=M^2$ and $M_-^2=0$.  Interestingly, the amplitude of these corrections no longer depends on $a_\phi$ but is entirely determined by the diagonal elements of the matrix $A$. 

There are several cases of interest in the resulting expression for $\gamma_{\rm eff}$. For short range fields, $Mr\gg 1$ in laboratory and solar system scales, the exponential term rapidly vanishes and we get 
\begin{equation}
\gamma_{\text{eff}}\approx \frac{3M^2-2m_\lambda^2}{3M^2+2m_\lambda^2}=1-\frac{4m_\lambda^2}{5m_\lambda^2+3m_\phi^2} \ .
\end{equation}
In order to have compatibility with current observations, we must have $|\gamma-1|<10^{-5}$, which implies that $m_\phi^2\ge 10^5 m_\lambda^2$. 
In the opposite extreme, we have the case of long range fields, $0<M r\ll 1$ over astrophysical scales, and leads to 
\begin{equation}
\gamma_{\text{eff}}\approx \frac{1}{5} \ ,
\end{equation}
which is in clear conflict with observations. The case $m_\lambda^2=m_\phi^2$ leads to important simplifications, 
\begin{equation}
\gamma_{\text{eff}}=\frac{2-e^{-\sqrt{2}m_\phi r}}{5+e^{-\sqrt{2}m_\phi r}} \ ,
\end{equation}
but does not improve in any way the viability of the theory, which is in clear conflict with observations. 

\subsection{Particular case $a_\phi=0$}

The particular case discussed above led to a partial decoupling between the scalar degrees of freedom of the general case, in the sense that the effective Newton constant and PPN parameter $\gamma$ did not depend on the parameters $a_\phi$ and $a_\lambda$, which are responsible for the direct coupling between  $\delta \tilde\phi$ and $\delta \tilde\lambda$ in Eqs. (\ref{eqphipert}) and (\ref{eqlambdapert}). A more obvious way to partially decouple these two degrees of freedom is by considering a situation with $a_\phi=0$, in such a way that the weak field dynamics of $\delta \tilde\phi$ becomes independent of $\delta \tilde\lambda$. This choice of $a_\phi$ constraints the effective potential to take the form
\begin{equation}\label{specialpot}
\tilde V(\tilde \phi,\tilde \lambda)=A(\tilde \phi)+B(\tilde \lambda)e^{2\sqrt{\frac{2}{3}}\kappa\tilde \phi}
\end{equation}
Here we discuss this particular case in some detail.  

Proceeding similarly as above, in this case Eq. (\ref{eqphipert}) decouples from Eq. (\ref{eqlambdapert}) and leads to the Fourier relation 
\begin{equation}
\delta\hat{\tilde \phi}=\sqrt{\frac{2}{3}}\frac{\kappa\hat\rho}{(k^2+m_\phi^2)} \ ,
\end{equation}
which can be inverted to obtain 
\begin{equation}
\delta\tilde\phi(r)=\sqrt{\frac{2}{3}}\frac{\kappa M_\odot}{4\pi r} e^{-m_\phi r} \ ,
\end{equation}
where again we assumed $\hat \rho(\vec{x}')=M_\odot \delta^{(3)}(\vec{x}')$. The Fourier modes corresponding to $\delta\tilde \lambda$ take the form
\begin{equation}
\delta\hat{\tilde \lambda}=-\frac{a_\lambda \delta\hat{\tilde \phi}}{(k^2+m_\lambda^2)} \ ,
\end{equation}
and after some algebraic manipulations we find its position space representation as
\begin{equation}
\delta\tilde\lambda(r)=-a_\lambda\sqrt{\frac{2}{3}}\frac{\kappa M_\odot}{8\pi } e^{-m_\phi r} \ ,
\end{equation}
 which has no $1/r$ behavior and, therefore, is finite at $r\to 0$ and decays at a much slower pace as $r\to \infty$. 

Finally, proceeding as in previous sections, the metric perturbations become
\begin{eqnarray}
h_{00}&=&\frac{2GM_\odot}{r}\left(1+\frac{2}{3}e^{-m_\phi r}\right), \\
h_{ij}&=&\frac{2GM_\odot}{r}\left(1-\frac{2}{3}e^{-m_\phi r}\right)\delta_{ij} \ ,
\end{eqnarray}
from which we extract
\begin{eqnarray}
G_{\text{eff}}&=&G\left(1+\frac{2}{3}e^{-m_\phi r}\right) , \\ 
\gamma_{\text{eff}}&=&\frac{3-2e^{-m_\phi r}}{3+2e^{-m_\phi r}} \ .
\end{eqnarray}
We readily see that, as expected, there is no trace of the scalar $\delta \tilde\lambda$ in these expressions, which has completely decoupled from the weak field limit. 
This case is also free from the repulsive Yukawa correction of the general case and also lacks of any constant shift associated to a zero mass mode.   
Obviously, only when $m_\phi r\gg 1$ will the theory pass the weak field observational tests. The situation is thus similar to what we found above in Sec. \ref{sec:C} but without any possibility to set bounds on the parameter $m_\lambda^2$ that characterizes the second scalar field at this perturbation level.

\section{Conclusions}\label{sec:concl}

We have studied the weak field, slow motion limit of hybrid metric-Palatini $f(R,{\cal R})$ gravity working in the Einstein frame of the corresponding scalar-tensor representation of this family of gravity theories. We have seen that the resulting dynamics is described by a metric and two dynamical scalar degrees of freedom, with the scalars mixing in different ways to yield a variety of scenarios. The results found here are consistent with those obtained by other means in \cite{Bombacigno:2019did}, though we identify various particular cases of interest not explicitly addressed in that work. This is, in part, possible thanks to the simplifications that our notation allows in the transit from the first line of (\ref{eq:deltaphi}) to the second line.
We have shown that, in the general case, the effective Newton constant is affected by both an attractive and a repulsive contribution, though the origin of the repulsive mode cannot be easily traced back to the negative sign with which one of the kinetic terms contributes to the total action. This is so because the only term that has a dependence on that sign, the constant $a_\lambda$, appears non-linearly in the effective parameters (via the quantity $M_0^2$ defined in Eq. \eqref{eq:M0}) and contributes in the same way to the amplitude of the Yukawa terms. The case of short range scalars and when the ratio $a_\phi/M^2_0$ are compatible with observations, while a scenario with long range fields cannot be ruled out, though it is harder to constrain.
We mention that we restricted our analysis of the general case to those cases in which the parameter $M_0^2$ is positive or zero. A negative value for this quantity would lead to oscillatory terms in the effective metric instead of the standard Yukawa-type corrections. Since there is no evidence supporting that kind of behavior, we omitted their discussion for the sake of clarity.

Furthermore, we pointed out the existence of two singular cases in the general discussion, namely, when $a_\phi=0$ and when $m_\phi^2 m_\lambda^2-a_\phi a_\lambda=0$, and analyzed them separately. In the latter case, we observed a partial decoupling of the scalar field $\delta\tilde\lambda$ from the weak field limit, whereas in the former this scalar is completely decoupled. We managed to establish some viability criteria for the $m_\phi^2 m_\lambda^2-a_\phi a_\lambda=0$ case, finding that one of the scalars must be much heavier than the other ($m_\phi^2>10^5 m_\lambda^2$), being short ranged. A similar requirement is needed in the $a_\phi=0$ configuration, though in this case there are no constraints on the mass $m_\lambda^2$. Note also that the decay of $\delta\tilde\lambda$ with the distance to the source is much slower than that of $\delta\tilde\phi$. Whether this may lead to relevant cosmological effects will be explored in more detail elsewhere.

\begin{acknowledgments}
JLR was supported by the European Regional Development Fund and the programme Mobilitas Pluss (MOBJD647).
FSNL acknowledges support from the Funda\c{c}\~{a}o para a Ci\^{e}ncia e a Tecnologia (FCT) Scientific Employment Stimulus contract with reference CEECINST/00032/2018, and thanks funding from the research grants No. UID/FIS/04434/2020, No. PTDC/FIS-OUT/29048/2017 and No. CERN/FIS-PAR/0037/2019.
GJO is funded by the Spanish projects FIS2017-84440-C2-1-P (MINECO/FEDER, EU), PROMETEO/2020/079 (Generalitat Valenciana), i-COOPB20462 (CSIC), and the Edital 006/2018 PRONEX (FAPESQ-PB/CNPQ, Brazil, Grant 0015/2019). 
The authors thank F. Bombacigno for useful comments. 
\end{acknowledgments}



\end{document}